# Experimental polarization encoded quantum key distribution over optical fibres with real-time continuous birefringence compensation

**G B Xavier[1,3], N Walenta[2], G Vilela de Faria[1], G P Temporão[1], N Gisin[2], H Zbinden[2] and J P von der Weid[1].**

[1] Centre for Telecommunication Studies, Pontifical Catholic University of Rio de Janeiro – R. Marquês de São Vicente 225 Gávea, Rio de Janeiro – Brazil

[2] GAP-Optique, University of Geneva, rue de l'Ecole-de-M´edecine 20, CH-1211 Geneva 4, Switzerland

E-mail: guix@opto.cetuc.puc-rio.br

**Abstract** We demonstrate an active polarization drift compensation scheme for optical fibres employed in a quantum key distribution experiment with polarization encoded qubits. The quantum signals are wavelength multiplexed in one fibre along with two classical optical side channels that provide the control information for the polarization compensation scheme. This setup allows us to continuously track any polarization change without the need to interrupt the key exchange. The results obtained show that fast polarization rotations of the order of 40 $\pi$ rad/s are effectively compensated. We demonstrate that our setup allows continuous quantum key distribution even in a fibre stressed by random polarization fluctuations. Our results pave the way for Bell-state measurements using only linear optics with parties separated by long distance optical fibres.

[3] Author to whom any correspondence should be addressed.

## 1. Introduction

Quantum Key Distribution (QKD) [1] was first experimentally demonstrated with polarization encoded photons sent over a free-space distance of 30 cm [2]. Shortly after, QKD experiments with polarization coding in optical fibres were performed [3]. However, phase coding [4] soon became the dominant form of transmission in guided media [5, 6], with polarization coding becoming the standard for free-space QKD links [7, 8].

A major problem of polarization coding in optical fibres arises from a randomly varying birefringence. It is a well-known phenomenon in classical optical communications that residual fibre birefringence modifies the State of Polarization (SOP) of a signal propagating along the fibre. Random fluctuations of this residual birefringence, caused by environmental changes, lead to an unpredictable time varying SOP at the receiver end. This, by itself, is not a major concern in classical telecommunications since most detection systems are insensitive to polarization fluctuations. However, such fluctuations are a concern in experiments which employ coherent detection and local laser oscillators. If the rotations were fixed, one could simply use a manual polarization controller to undo the transformations caused by birefringence. But, due to the random nature of the fluctuations, active polarization tracking devices [9] or polarization diversity schemes [10] must be used in order to circumvent this limitation for polarization-sensitive detection schemes in optical communications. For classical communications this problem is solved, as only a single SOP must be tracked and part of the signal itself can be used as feedback for the real time tracking device. For quantum communication this problem remains a challenge since the transmitted polarization states are non-orthogonal. On the other hand, the signal itself cannot be used for feedback to the polarization tracking device due to its quantum nature.

Recently, active polarization control schemes for quantum communication have been successfully implemented to distribute polarization encoded qubits over optical fibres [11, 12]. In both cases, however, the QKD session was interrupted at times to perform the polarization tracking. Thus, compensation was restricted only to polarization variations slow enough to allow key transmission without control for a while. As QKD rates and reliability requirements increase, on line control schemes should be developed to reach these demands.

A different polarization based QKD scheme employing a fibre link was proposed in [13, 14]. In this approach, entangled photon pairs with orthogonal polarizations were sent from Alice to Bob, dismissing the use of a shared reference frame for the polarization states. But the requirement to detect both photons of a pair imposes more severe restrictions on the link attenuation as well as on the detection apparatus than those normally used in typical polarization based QKD experiments.

A real time polarization control scheme, able to track *all* polarization states at the receiver end is of great interest not only for QKD but also for other applications, such as partial long-distance Bell state measurements employing linear optics [15] and quantum relays and repeaters [16, 17]. Such a control scheme was recently realized using two ITU-T (International Telecommunication Union - Telecommunication Standardization Sector) standard grid side channels adjacent to the quantum signal wavelength in order to provide the necessary classical feedback to the control loop [18].

In this paper we present, to the best of our knowledge, the first quantum communication experiment with polarization encoding in optical fibres with true real-time continuous polarization control. To demonstrate the robustness of the scheme we employed a polarization scrambler to further randomize the unitary rotation transformations performed by the fibre. The whole experiment is performed using only off-the-shelf standard telecom components, making it a practical implementation to be used in future polarization encoded quantum communication experiments.

## 2. Control theory

The relationship between the output and input quantum polarization states in an optical fibre can be written as:

$$|\Psi\rangle_{OUT} = U_F |\Psi\rangle_{IN} \tag{1}$$

where $U_F$ is the unitary operator representing the rotations caused by random birefringence changes in the fibre. Controlling *all* polarization states over the fibre means that one should be able to undo the unitary transformation $U_F$. Therefore, the control system must be able to perform a unitary transformation $U_T$, where $U_T = U_F^{-1}$ such that the output quantum state is $|\Psi\rangle_{OUT} = U_T U_F |\Psi\rangle_{IN} = |\Psi_{IN}\rangle$.

The system used in this paper employs two classical reference signals, henceforth called $|S_1\rangle$ and $|S_3\rangle$ that will carry the control information needed to perform $U_T = U_F^{-1}$. As it has been recently shown, full polarization control is achievable if two non-orthogonal reference SOPs $|S_1\rangle$ and $|S_3\rangle$ are launched in the fibre together with the signal to be controlled [18]. At the end of the fibre a polarization controller is used that performs two rotations $R_1$ and $R_3$ in series. Bound by the polarization controller the following set of equations holds:

$$\begin{cases} R_1 U_F |S_1\rangle = |S_1\rangle \\ R_3 |S_1\rangle = |S_1\rangle \\ R_3 R_1 U_F |S_3\rangle = |S_3\rangle \end{cases} \tag{2}$$

$R_1$ and $R_3$ cancel out the rotations caused by the fibre on the states $|S_1\rangle$ and $|S_3\rangle$. We note that $R_3$ does not have any effect on $|S_1\rangle$, as it is a rotation around the axis formed by $|S_1\rangle$ and its orthogonal state on the Poincaré sphere. By rewriting (2) in matrix form we can conclude that $R_3 R_1 U_F$ must be equal to the identity, that is, $R_3 R_1 = U_F^{-1}$.

In the above discussion $U_F$ is considered to be wavelength independent. Since fibre birefringence varies for different wavelengths, perfect polarization control would only be possible if $|S_1\rangle$ and $|S_3\rangle$ have the same wavelength as the signal we would like to control. By including the wavelength dependency, *group refractive indexes* instead of *phase refractive indexes* are concerned and the evolution of the SOP is characterized by the group delay between the polarization modes. As the differential group delay is randomly varying in time, the fibre is then characterized by its mean group delay value $<\tau>$, usually of the order of a fraction of a picosecond for a typical 100-km fibre. We can still achieve good control performance as long as both the wavelength separation and the fibre mean group delay are sufficiently small, $\tau \Delta\omega << 1$. Good results were achieved even with a channel separation of 0.8 nm and a fibre mean group delay of 0.54 ps [18]. Simulations have shown that good control with the same channel separation on fibres with $<\tau>$ up to 1 ps are achievable [19], meaning that a good correlation exists between the SOP of the reference signals and of the quantum channel. Most QKD fibre spans (~100 km) have $<\tau>$ smaller than 1 ps.

## 3. Control in practice

In an experimental implementation additional measures must be taken in order to employ the polarization control scheme. As just mentioned, classical signals containing the non-orthogonal polarization states $|S_1\rangle$ and $|S_3\rangle$, to be used in the feedback control loop, must be present in the fibre with the single photons. Their intensities normally depend on the fibre distance, but are typically orders of magnitude higher than the single photons, which would make any quantum transmission impossible due to noise levels. Therefore, the reference signals must be time or wavelength multiplexed with the single photons.

The simplest solution is to send the reference information in the time intervals between quantum signals, and tuned on the same wavelength. Since this approach works by having all signals on the same wavelength, it works for any transmission distance, as the polarization correlation between channels is always high. One drawback is that (in practice) it is incompatible with high quantum data rates, as the control scheme will have difficulties stabilising the polarization in the short time span in between single photons. Experiments were performed using this technique, but the control was not used in between every single qubit sent. Instead the polarization changes were compensated every so often, and during this time, the quantum transmission was completely stopped, lowering the overall transmission rate [11, 12].

Our proposal consists of a real-time continuous solution where the reference signals are separated in wavelength from the quantum signal, in order to achieve this goal. As was discussed in [19], there are three possible approaches: a) one control channel, containing both SOPs $|S_1\rangle$ and $|S_3\rangle$, each with a different amplitude modulation frequency, easily split at the receiver through filtering (dithering); b) two separate control channels (one for each SOP) operating in continuous-wave mode (CW); and c) combining the solution from a) and b), that is, two separate control channels, each containing both states $|S_1\rangle$ and $|S_3\rangle$ through amplitude modulation, so that the mean value is taken. Method c) yields the best control performance, although it is more complex to implement. In this paper method b) is used, as it is simpler to implement, and it gives the same performance as method a) [19].

The choice of wavelengths for the control channels is not arbitrary. The correlation between polarization fluctuations from different wavelengths depends strongly on the separation between channels. One expects that the closer the wavelengths the stronger the correlation, and our simulations have indeed confirmed that this is the case [19]. For the scheme used in this work we employ two side channels for the feedback signals, one at a longer wavelength than the single photons and the other at a shorter one. Since we want them to be as close as possible, and we would like to be able to use standard off-the-shelf telecom components for an easier implementation, we use a wavelength separation of 0.8 nm (100 GHz around 1550 nm), which is the standard value for Dense Wavelength Division Multiplexing (DWDM) systems employed in optical networks following the ITU-T grid (Figure 1).

Adequate filtering must be provided in order to implement our scheme, as the classical reference channels have much higher intensities than the single-photon levels of coherent states used in quantum communications with attenuated pulsed lasers. Any photons generated from the reference channel lasers and detected by the Single Photon Counting Modules (SPCMs) are perceived as noise, degrading the transmission, or even stopping it completely. We can isolate two noise contributions in this scheme: the first is caused by “real” components, such as filters with finite extinction ratios and lasers emitting light outside of their centre wavelength. The second contribution comes from the fibre itself, in the form of non-linear and scattering effects [20].

The first contribution can be removed with careful filtering and can stem from Amplified Spontaneous Emission (ASE) generated from a laser. ASE noise extends far beyond the laser centre wavelength, generating broadband noise several tens of nanometres wide. ASE noise from lasers is not, in most cases, a concern for classical optical telecom systems, since its power is usually around 40 dB below the centre wavelength from the laser. When working in the single-photon counting regime, ASE optical power levels falling on the SPCM are more than enough to completely blind the detector. This issue can be easily circumvented with careful filtering of the reference laser, to remove the ASE noise and only allow the centre wavelength to be multiplexed with the classical side-channel. We perform this filtering with Fibre Bragg Gratings (FBGs), which are explained in more detail in the next section. DWDMs provide the next layer of filtering to properly multiplex and demultiplex the classical and quantum channels, and to minimize crosstalk. The same filtering configuration also removes noise from Rayleigh backscattering generated in the fibre, since the scattered photons have the same wavelength as the original ones.

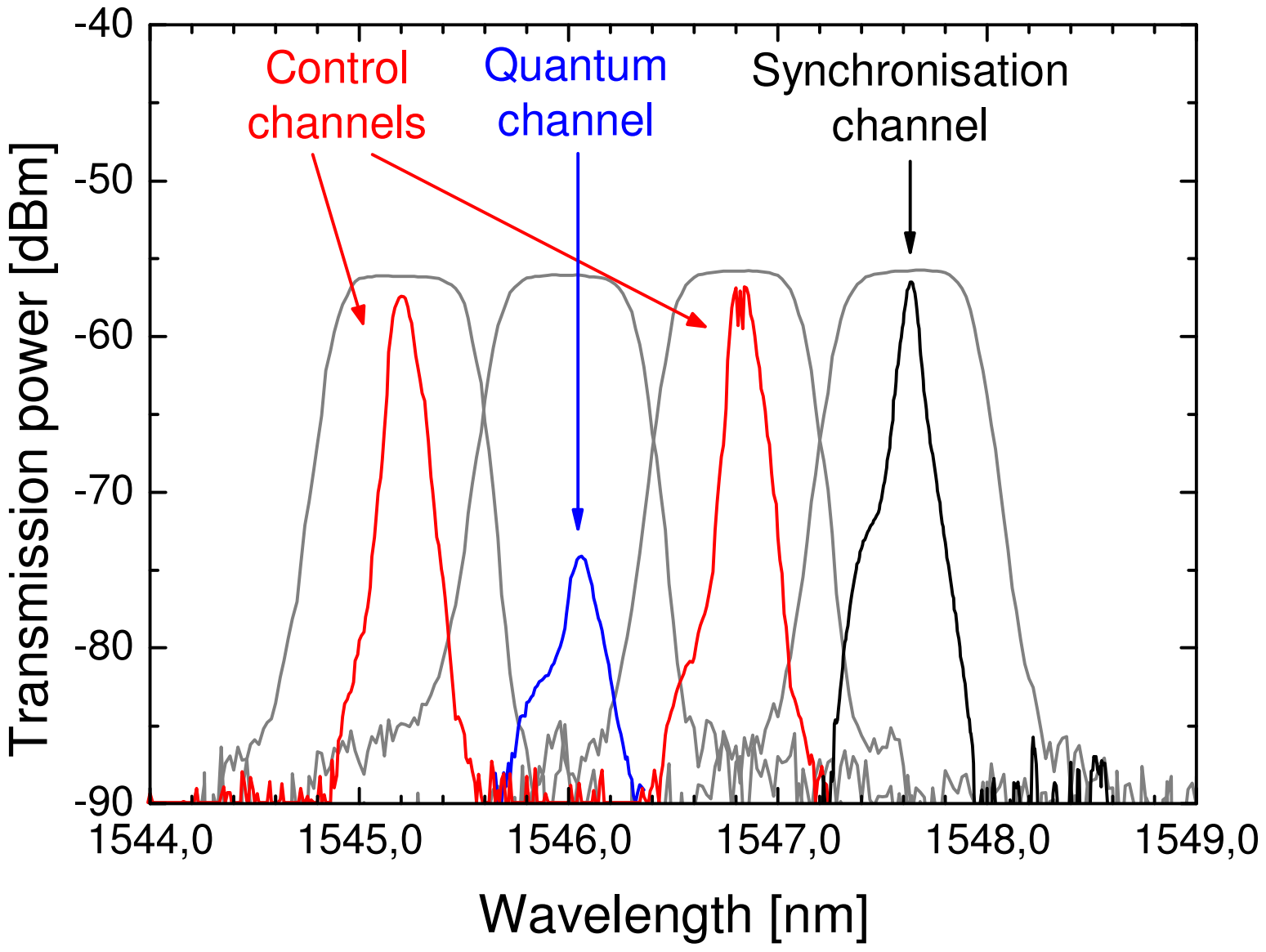

**Figure 1**. Emission spectra of the two polarization control lasers (red), quantum channel laser (blue) and synchronisation laser (black) aligned to 4 adjacent channels of the ITU-C band between 1545.32 – 1547.72 nm. The transmission spectra of the respective DWDM channels are shown as grey lines. The lasers spectra were measured after the Fiber Bragg Gratings (FBG).

As mentioned above, another noise contribution comes in the form of non-linear and scattering effects generated inside the fibre. One known problem of simultaneously transmitting wavelength separated classical channels alongside a quantum signal inside an optical fibre is that of spontaneous Raman scattering [21, 22]. In principle, it cannot be completely removed by filtering since a considerable fraction of the photons generated over the fibre's length have the same wavelength as the quantum channel. Fortunately for our setup, Raman induced noise from a neighbouring channel is smaller when the wavelength separation between a classical signal and a quantum channel is the smallest possible (Figure 2) [23]. Still, this is not enough to remove all the noise, since the Raman induced count probability on the detector is still quite high.

The solution we take to remove the residual Raman noise is to create a short dark slot in which the pseudo-single photon pulse is transmitted. This approach is taken since, as shown in [21, 22, 23], the noise we detect is Raman *spontaneous* induced noise, which varies linearly with input power, unlike Raman *stimulated* scattering which is non-linear in nature. Since it is a linear contribution, it is harder to remove it by simply attenuating the side channels. As explained below our solution is effective and it does not compromise the control performance.

Another method to reduce spontaneous Raman scattering would be to use narrower filters on the quantum channel since the noise is broadband. There are commercially available fibre based filters composed of Bragg gratings with a bandwidth of 10 pm [24]. Since the measurements shown in figure 2 used a 0.3 nm wide filter, we can expect a reduction of a factor 30 in the noise by using the narrower filter. A similar reduction can be achieved by using shorter detection gates, and they should get narrower as detector technology evolves. Since the amount of noise increases with the fibre length, shorter fibre spans will also have a lower noise contribution. However, it will only be negligible when its length is very short.

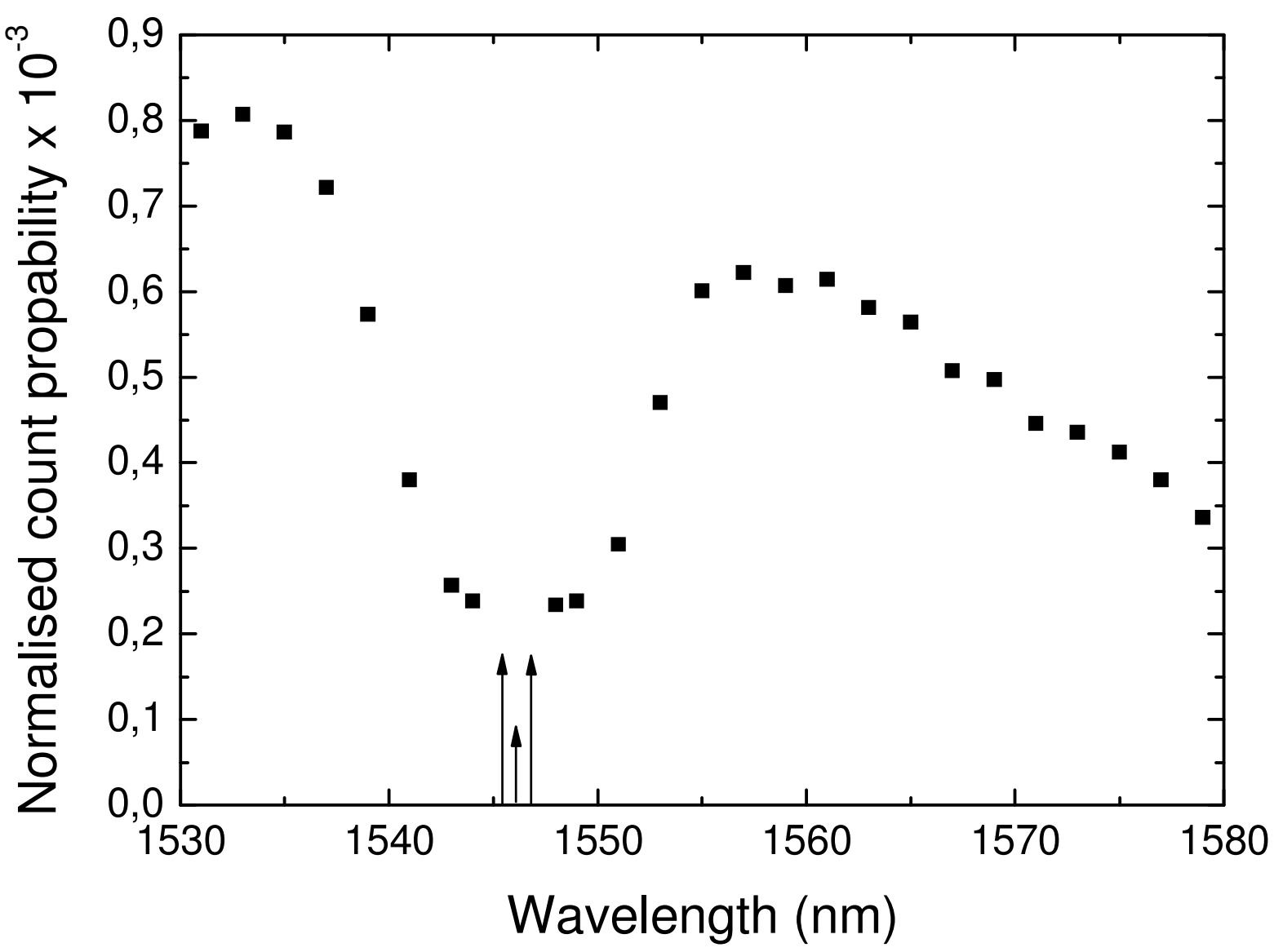

**Figure 2.** Measured count probability per detection window on an InGaAs commercial single-photon detector normalized to a 1 ns gate window over 8 km of dispersion-shifted optical fibre. The wavelength sweep is performed with a 1 mW CW tuneable laser co-propagating with the transmission direction of the single photons. The detector has a fixed filter centred at 1546.12 nm (the quantum channel wavelength to be used in our scheme). The three arrows at the bottom represent the wavelengths our system will work on. Note that the count probability is considerably higher than the typical dark count probability per gate window on commercial detectors ($\sim 10^{-5}$). It was verified through appropriate filtering that the noise counts were generated by the spontaneous Raman effect inside the fibre.

## 4. Experimental demonstration of a polarization encoded QKD session

The experiment we perform here is schematically presented in figure 3. It employs two classical control side channels generated from standard telecom DFB (Distributed Feedback) lasers centred at $\lambda_1 = 1545.32$ nm and $\lambda_3 = 1546.92$ nm respectively. The quantum channel uses a pulsed attenuated DFB laser centred at $\lambda_Q = 1546.12$ nm. All three channels are multiplexed at Alice's setup using a commercial DWDM multiplexer with 1.4 dB insertion loss and an extinction ratio of at least -35 dB between adjacent channels and -45 dB between non-adjacent ones. Each side channel employs a circulator with a FBG centred at $\lambda_Q$ to remove any ASE noise generated from the respective lasers, which would fall on the SPCM at Bob's side, rendering QKD impractical.

Our control system is designed to work with the reference lasers working in CW mode. However, as just discussed, spontaneous Raman scattering induced noise will cause false detections on the SPCM. To avoid this problem, when we send the quantum signal we reduce the power of the side channel lasers to less then -90 dBm by modulating the current, thus creating a dark temporal slot of 13.5 ns, yielding a duty cycle for the control channels of 0.93, corresponding to the transmission frequency of our setup of 5 MHz. This modulation is transparent to our control scheme due to an electronic low-pass filter on the detection side and does not affect the overall performance. Through this method we can guarantee a continuous control that works full-time and does not degrade the performance of the quantum communication taking place.

For the experiments presented here we use the electronics that controlled a "plug-and-play" QKD phase coding setup [5] and adapted the optical part for our polarization-based scheme. To temporally synchronize with Alice, Bob generates 1 ns long pulses with a repetition frequency of 5 MHz. These clock signals are transmitted in trains of 300 pulses in the same fibre as the quantum and classical side channels to Alice using a DFB laser centred at $\lambda_S$ = 1547.72 nm. In principle, backscattered photons from the synchronisation pulses would induce noise in the quantum channel due to Raman and crosstalk from Rayleigh backscattering. However, these effects are not relevant in our setup since, at Alice, we delay the incoming signals by 50 µs in order to gain time to synchronize her internal clock with the incoming pulses before triggering the emission of the quantum signals. This delay acts as storage for the synchronisation pulses such that no intersection of quantum signals and backscattered light takes place in the transmission fibre. We verified that the synchronisation pulses did not affect the noise on the SPCMs at Bob.

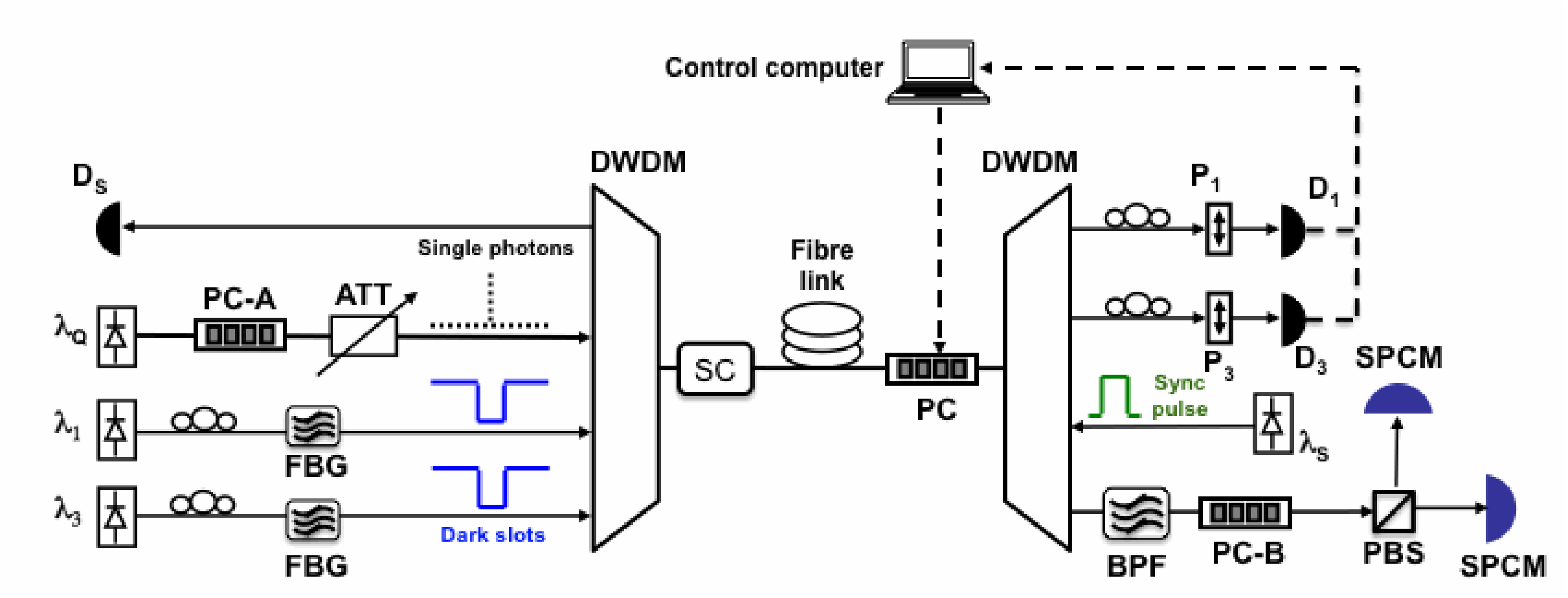

**Figure 3.** Experimental setup. Solid line represent optical fibres while dashed ones account for electrical cables. $D_S$: Synchronisation classical detector; ATT: Optical attenuator; $D_1$ and $D_3$: Side-channels classical detectors; FBG: Circulator and fibre Bragg grating; PC, PC-A and PC-B: Control system, Alice and Bob's $LiNbO_3$ polarization controllers respectively; DWDM: Dense wavelength division multiplexer; $P_1$ and $P_3$: Linear polarizers; SC: Fibre polarization scrambler; SPCM: Single photon counting module; BPF: Band-pass filter, which is an identical DWDM as the ones used by Alice and Bob. The direction of the propagating pulses is indicated on the figure. The electronic circuitry used to modulate the side-channel laser currents to create the dark slots of 13.5 ns mentioned in the text above, as well as all delay generators used to synchronize signals are omitted from the figure for the sake of clarity.

The SOP of the faint laser pulses is modified with a fast $LiNbO_3$ fibre pig-tailed electro-optic polarization controller (PC-A, EOSPACE), with the modulating electrical signal generated from a FPGA (Field Programmable Gate Array) passing through a high voltage electronic driver ($V_\pi$ ~ 50 V). This controller switches between the four distinct SOPs needed in the BB84 protocol [1, 2]. An identical set of $LiNbO_3$ controller and driver is used by Bob to change between the two measurement bases (PC-B). The modulator in our setup was able to change between orthogonal polarization states within 10 ns, as shown in figure 4. This measurement was performed by modulating the polarization state of a CW laser with one of the $LiNbO_3$ controllers, passing the optical signal through a polarizer and measuring the signal intensity with a p-i-n photodiode. From this we verify that our modulation speed is compatible with the 5 MHz repetition frequency generated from the electronics.

For the link between Alice and Bob we employed a 16 km spool of SMF-28 optical fibre with 4.3 dB losses and 0.28 ps mean differential group delay. Before the fibre link we added a polarization scrambler, composed of three separate piezo-electric actuators, to demonstrate the effectiveness of our system even in the presence of fast polarization changes. An identical $LiNbO_3$ polarization controller (PC) as the ones used by Bob and Alice, placed after the fibre spool, is driven by the control computer and performs the inverse unitary transformation $U_T = U_F^{-1}$ on all channels. The $LiNbO_3$ polarization controller cannot endlessly rotate the polarization state of the light. However, since it is composed of several wave-plates (six in our case) oriented at alternating angles (0º, 45º, 0º,…) there is a considerable degree of redundancy. This means that every time a one of them has reached its rotation limit the next one is ready to take over. If all wave-plates reach their rotation limit at the same time the control system resets the polarization controller, which would cause errors in the transmission at a short period of time. As such an event is highly unlikely in our case due to the controller redundancy, we did not observe such a situation in our experiments.

The optical signal at Bob is demultiplexed with an identical DWDM as the one used by Alice. To decrease noise from crosstalk we further increased the extinction by filtering the quantum channel with a second DWDM (band-pass filter BPF on figure 3). The quantum signal passes Bob's $LiNbO_3$ polarization controller (PC-B) which performs the basis choices before they are analysed at a polarising beam splitter (PBS) and detected by the SPCMs.

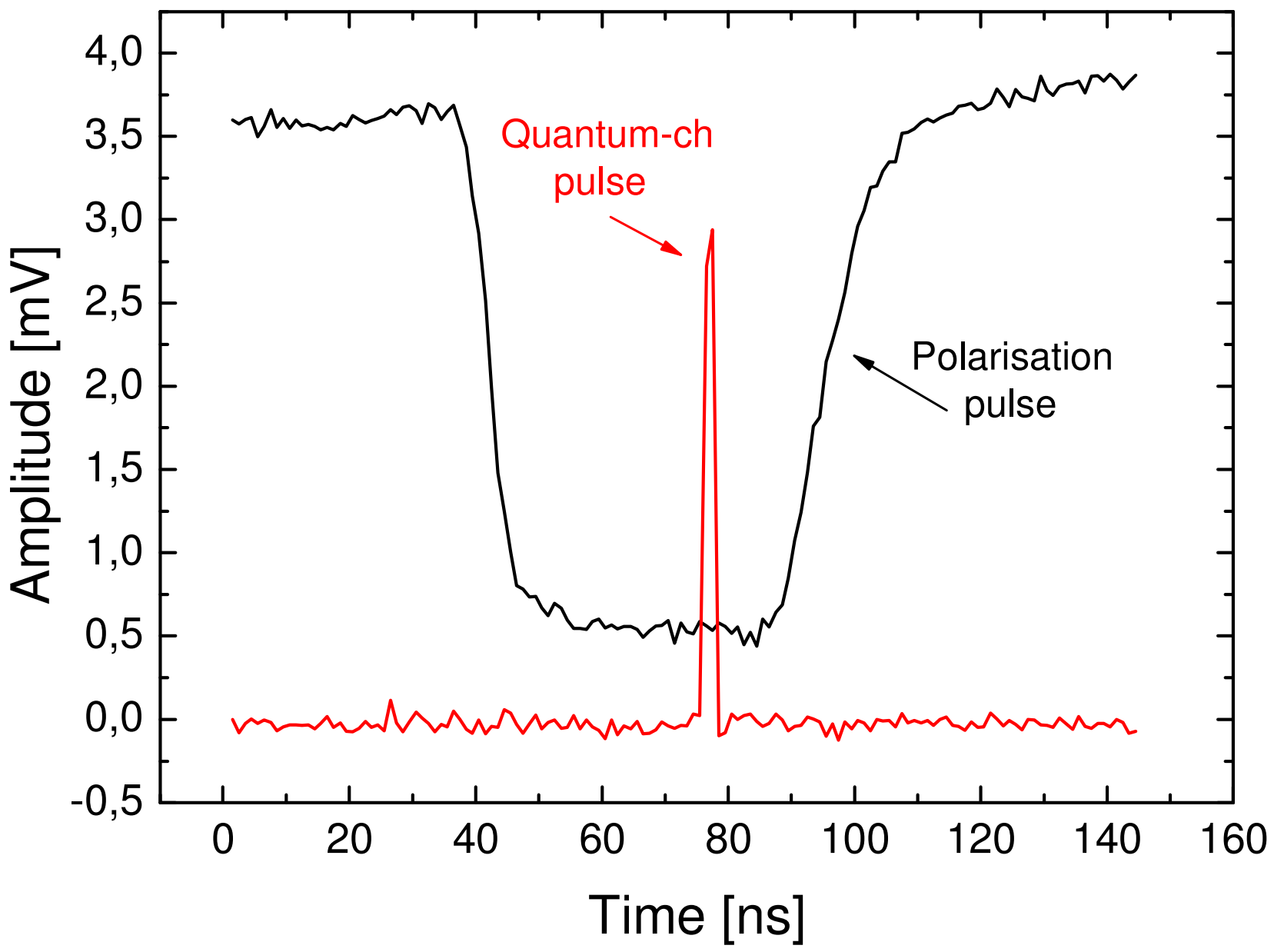

**Figure 4**. Intensity measurements of a polarization pulse (black), and the quantum channel signal (red), operating on classical power levels. The polarization pulse was taken after a polarizer, with a CW laser, while switching the SOP between two orthogonal values, and back to the original. The quantum channel pulse is included here as a reference, showing that it is much narrower than the polarization bit.

The classical side channels are separated by the DWDM and pass through linear polarizers with their optical axis adjusted to be non-orthogonal using manual polarization controllers [18]. Their optical intensities after the polarizers are measured by classical p-i-n photodiodes and fed to the control computer for processing. Its control algorithm maximizes the intensity of both control channels at the same time by changing the polarization state of the optical signals before splitting at the DWDM. By maximizing both side channel intensities simultaneously, the original input polarization states are recovered.

The whole setup is controlled by personal computers that send out the necessary electrical signals to the optical components to perform a QKD session, and perform the classical procedures required by BB84. The classical channel between Alice and Bob was realized with an USB connection between both systems.

Before performing a quantum transmission, the entire setup is calibrated by adjusting the polarization of the quantum channel as well as both side channels. For the quantum channel, two manual polarization controllers (not shown in figure 3) before PC-A and PC-B were used to align the input polarization state with the $LiNbO_3$ polarization modulator for identical maximum rotations on the Poincaré sphere. Another manual polarization controller (also not shown in figure 3) is placed at Bob after his PC-B to align the state to the axis of the subsequent polarizer.

The polarization states of the side channel lasers were set with two manual controllers to be non-orthogonal. We have improved on the previous scheme in such a way that the control will work properly as long as the side-channels are non-orthogonal, but not necessarily maximally overlapped as it used to be the case [18], which makes the system more robust and considerably easier to align than before. The control only fails to compensate the polarization drifts if the SOPs of the two control signals $|S_1\rangle$ and $|S_3\rangle$ are close to being situated on the same axis in the Poincaré sphere. The alignment procedure only needs to be done once, before initializing the transmission. We note that this adjustment could be performed automatically by employing additional $LiNbO_3$ controllers, and as such, our system could be used in a commercial environment where no manual intervention is needed.

After alignment of the entire system, we tested the performance of the setup, initially with the stabilisation system installed but not active. Alice's quantum signals were attenuated to obtain an average of $\mu = 0.1$ photons per pulse. With a side channel laser power of -7.4 dBm each, but without the stabilisation system running and the polarization scrambler turned off, the prepared states yielded a visibility of 97.2 % corresponding to a minimal QBER (Quantum Bit Error Rate) of 1.4 %. The measured QBER was 1.6 %, with an optical share $QBER_{opt} = 0.7$ % which is caused by the detection of photons in the wrong detector, mainly due to the limited 22 dB extinction of Bob's polarising beam splitter. Another share of $QBER_{det} = 0.1$ % is caused by noise counts due to the SPCMs and a share of $QBER_{side} = 0.8$ % is caused by noise due to crosstalk and photons generated by Raman scattering in the side channels. With the stabilisation system active, the QBER increased by 1.1 %. This can be ascribed to an increase in the optical share $QBER_{opt}$ due to fluctuations of the polarization state inherently induced by the stabilisation algorithm.

Before demonstrating QKD we characterize the performance of the stabilisation system even in the presence of fast polarization changes using the scrambler. A voltage ramp can be applied to each piezo crystal, which performs a polarization rotation of $2\pi$ back and forth on the Poincaré sphere at a tuneable frequency. As such, a scrambling frequency of 1 Hz means a polarization rotation of $4\pi$ per second. It should be noted that such extreme polarization fluctuations are rarely expected in normal environments.

In order to reduce the influence of detector and side channel noise to less than 0.1 % during the following characterisation, we increase the average photon number to $\mu = 1.0$ per pulse. We constantly prepare the same state at Alice and measure in the corresponding basis at Bob, in order to eliminate any possibility of errors induced by the polarization modulators. Figure 5 then shows the optical share $QBER_{opt}$ measured at different voltage ramp frequencies applied to the scrambler. Each point is averaged over 50 measurements, with 1 million photon pulses sent from Alice to Bob per measurement. The results show that $QBER_{opt}$ stays constantly under 6 % for scrambling frequencies up to 16 $\pi$/s, and increasing the rotations to 40 $\pi$/s, $QBER_{opt}$ has an average of 7.5 %.

To demonstrate the applicability of the stabilisation system for QKD under the condition of random polarization changes, as it occurs in aerial fibres or under thermal or mechanical stresses, we replaced the piezo-electric scrambler by a manual polarization modulator. We also decreased the average photon number per pulse to $\mu = 0.1$ to perform a secure transmission. Unfortunately, an electronic problem with Bob's polarization modulator PC-B reduced the stability and extinction of its modulation during a key exchange at 5 MHz. This forced us to simulate a random key exchange by measuring Alice's randomly prepared qubits first in one basis at Bob and then, in a subsequent measurement, in the other basis. In figure 6 both measurements are combined on top of each other with black points indicating key exchanges with measurements in Bob's first basis and red points in the second, respectively. This problem, along with errors in the alignment, is also the reason for the slightly increased QBER during the key exchange demonstration.

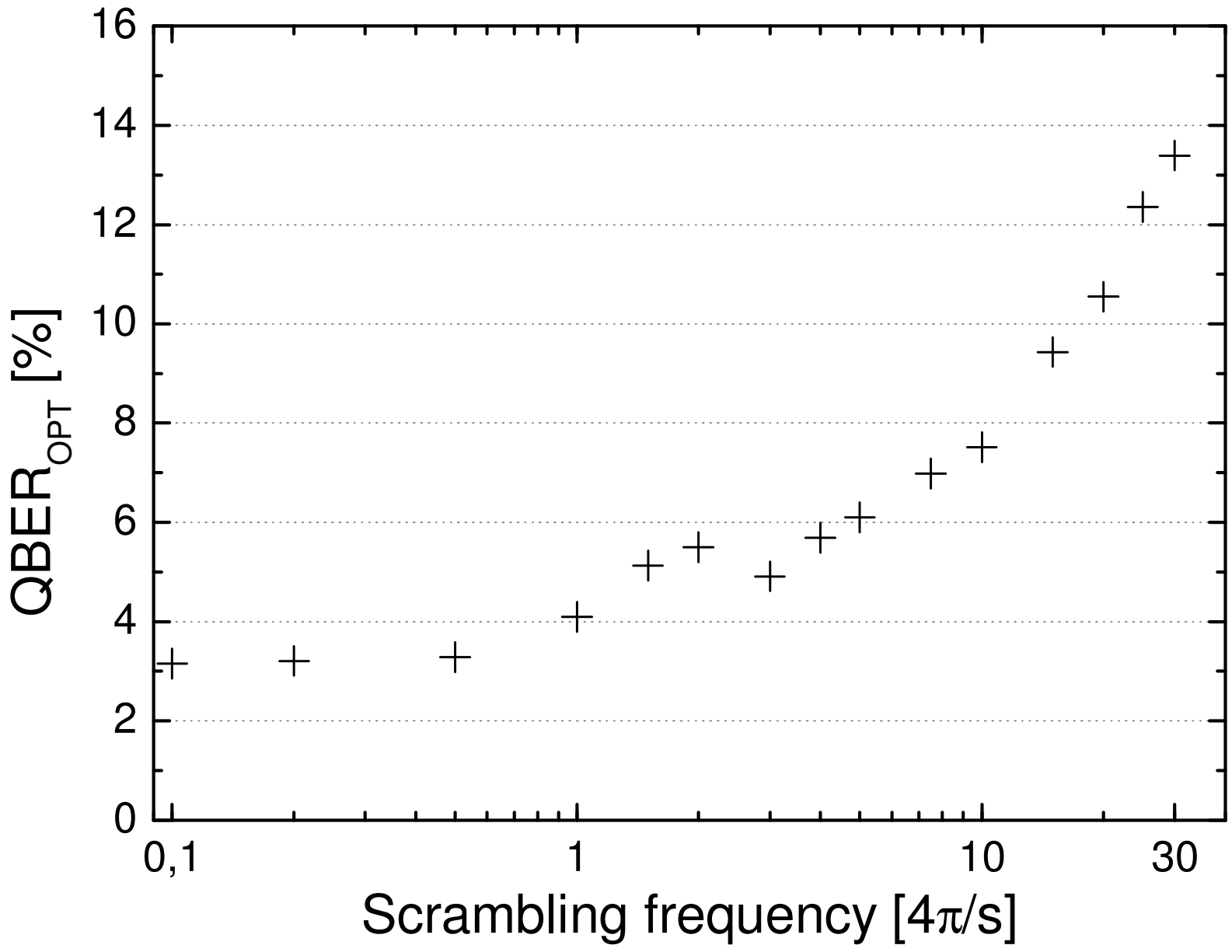

**Figure 5.** The optical share $QBER_{opt}$ as a function of the scrambling frequency demonstrating the stabilisation capability of the control system under rapid polarization changes. Each value is averaged over 50 measurements, with 1 million photon pulses sent per measurement.

Still, figure 6 shows the effectiveness of the stabilisation system when the polarization state is randomly scrambled in the order of a few rad/s. Again, each point represents a key exchange of one million sent qubits. In the first section (figure 6, a) keys are distributed with the stabilisation system running but without any polarization scrambling. In the second section (figure 6, b) the average QBER increases by only 1.2 % during the key exchange although the polarization states are continuously and randomly scrambled. By comparison, without the stabilisation system the QBER would increase dramatically with an average of ~ 50 % (figure 6, c) making any quantum key distribution impossible. The last section (figure 6, d) reveals that the system is able to re-stabilize immediately when it is reactivated.

## 5. Conclusions

The scheme implemented here demonstrates that it is possible to achieve real-time continuous control of the polarization state of single photons along a 16 km long optical fibre link, with an active polarization scrambler connected in series. We have demonstrated the feasibility of quantum key distribution employing polarization encoded qubits in optical fibres, in situations where the SOP of the transmitted photons is subject to fast random variations. The scheme was assembled using only standard off-the-shelf telecom components, and can be used with other single-photon sources, such as those based on spontaneous parametric down-conversion [25]. Furthermore our setup allows any quantum communication requiring polarization encoding in long-distance optical fibre links.

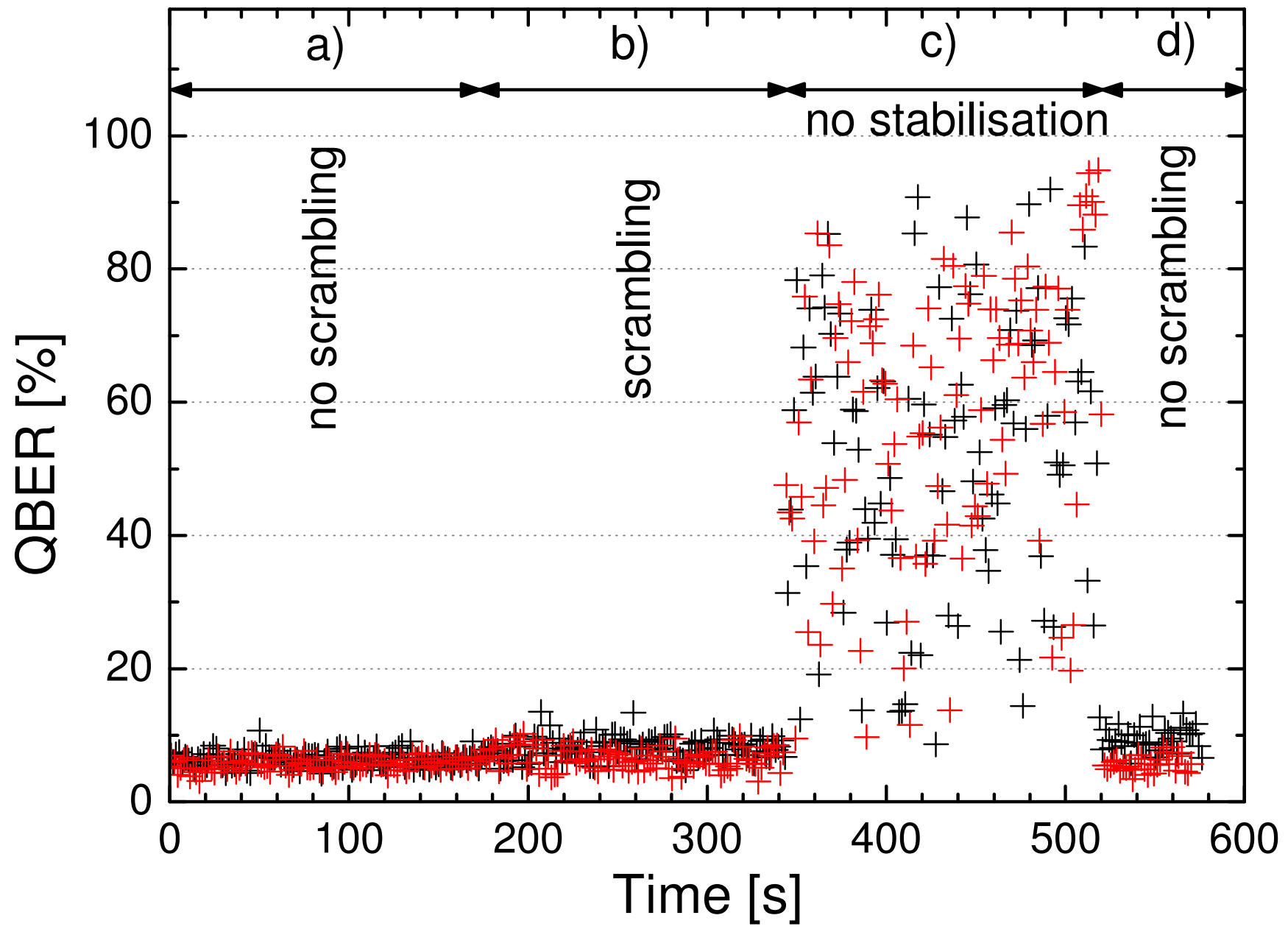

**Figure 6.** QBER as a function of time under different conditions. a) No polarization scrambling. b) Polarization scrambling with active stabilisation. c) Polarization scrambling without stabilisation system. d) Re-stabilization after the system is reactivated. Each point corresponds to 1 million sent qubits. Black and red points distinguish measurements in different bases at Bob (see text for details).

## Acknowledgements

The authors wish to thank FAPERJ, CAPES, CNPq and NCCR Quantum Photonics of the Swiss National Science Foundation for the financial support. We would also like to thank Marcelo Jimenez, Jean-Daniel Gautier, Claudio Barreiro, Patrick Eraerds and Rob Thew for their helpful contributions.

## References

[1] Gisin N, Ribordy G, Tittel W and Zbinden H 2002, *Rev. of Mod. Phys.* **74** 145.
[2] Bennett C H, Bessette F, Brassard G, Salvail L and Smolin J 1992, *J. Cryptology* **5** 3.
[3] Muller A, Breguet J and Gisin N 1994, *Europhys. Lett.* **23** 383.
[4] Townsend P D, Rarity J G and Tapster P R 1993, *Electron. Lett.* **29** 634.
[5] Stucki D, Gisin N, Guinnard O, Ribordy G and Zbinden H 2002, *New J. Phys.* **4** 41.
[6] Gobby C, Yuan Z L and Shields A J 2004, *Appl. Phys. Lett.* **84** 3762.
[7] Schmitt-Manderbach T, et al 2007, *Phys. Rev. Lett.* **98** 010504.
[8] Ling A, Peloso M P, Marcikic I, Scarani V, Lamas-Linares A and Kurtsiefer C 2008, *Phys. Rev.*

A **78** 020301(R).
[9] Walker N G and Walker G R 1990, *J. Lightwave Technol.* **8** 438-458.
[10] Kazovsky L G 1989, *J. Lightwave Technol.* **7** 279-292.
[11] Peng C-Z, et al 2007, *Phys. Rev. Lett.* **98** 010505.
[12] Chen J, Wu G, Li Y, E. Wu E, and H. Zeng H 2007, *Opt. Express* **15** 17928.
[13] Boileau J-C, Laflamme R, Laforest M and Myers C R 2004, *Phys. Rev. Lett.* **93** 220501.
[14] Chen T-Y, Zhang J, Boileau J-C, Jin X-M, Yang B, Zhang Q, Yang T, Laflamme R, and Pan J-W 2006, *Phys. Rev. Lett.* **96** 150504.
[15] Weinfurter H 1994, *Europhys. Lett.* **25**, 559.
[16] de Riedmatten H, Marcikic I, Tittel W, Zbinden H, Collins D and Gisin N 2004, *Phys. Rev. Lett.* **92** 047904.
[17] Yuan Z-S, Chen Y-A, Zhao B, Chen S, Schmiedmayer J and Pan J-W 2008, *Nature (London)*, **454** 1098.
[18] Xavier G B, Vilela de Faria G, Temporão G P and von der Weid J P 2008, *Opt. Express* **16** 1867.
[19] Vilela de Faria G, Ferreira J, Xavier G B, Temporão G P and von der Weid J P 2008, *Electron. Lett.* **44** 228
[20] Agrawal G P 2001, *Nonlinear fiber optics 3rd edition*, Academic Press San Diego USA.
[21] Imeke N I, et al 2004, *Appl. Phys. Lett.* **87** 174103.
[22] Subacius D, Zavriyev A and Trifonov A 2005, *Appl. Phys. Lett*. **86** 011103.
[23] Xavier G B, Vilela de Faria G, Temporão G P and von der Weid J P 2008, Proc. of QCMC 2008 Calgary.
[24] Halder M, Beveratos A, Gisin N, Scarani V, Simon C and Zbinden H 2007, *Nature Phys.* **3** 692.
[25] Halder M, Beveratos A, Thew R T, Jorel C, Zbinden H and Gisin N 2008 *New J. Phys.* 10 023027.